 \documentclass[%
 reprint,
citeautoscript
bibnotes,
 amsmath,amssymb,
 aps,
prb,
floatfix,
]{revtex4-2}

\usepackage{graphicx,rotating,subfigure,amsmath,amsfonts,amssymb,delarray,color}
\usepackage{hyperref}
\usepackage{xcolor}
\usepackage{soul}
\usepackage{dsfont}
\usepackage[T1]{fontenc}
\usepackage{physics}
\usepackage{multirow}
\usepackage{bbm}
\usepackage{dcolumn}
\usepackage{braket}
\usepackage{float}
\numberwithin{equation}{section}

\begin{document}

\title{Band Structures of One-Dimensional Periodic Materials with Graph Theory}
\author{R. Gerstner}
\affiliation{Department of Physics, McGill University, 3600 Rue University, Montréal, QC H3A 2T8}
\date{December 19, 2024}

\begin{abstract}
We show how arbitrary unit cells of one-dimensional periodic materials can be represented as graphs whose nodes represent atoms and whose weighted edges represent tunneling connections between atoms. Further, we present methods to calculate the band structure of a material with an arbitrary graphical representation, which allows one to study the Fermi level of the material as well as the conductivity at zero temperature. We present results for both circular chains as well as randomly-generated unit cell structures, and also use this representation to show that the connectivity of the unit cell is not correlated to its band gap at half filling. This paper provides an introductory insight into the utilization of graph theory for computational solid-state physics.
\end{abstract}

\maketitle

\section{Introduction}
\label{intro}

The field of solid state physics is characterized by the study of materials and the emergence of their properties from the atomic scale. The task is made difficult by the extremely large numbers of atoms in macroscopic systems, resulting in the the need for approximations or the exploitation of certain simplifying properties. A large class of materials, such as crystals and polymers, exhibit an important property called \textit{periodicity}, whereby the atoms and/or molecules are arranged in a repeating periodic pattern. This periodicity may occur in one, two, or three dimensions. Knowing a material is periodic can simplify the calculation of its properties by dividing the material into its smallest, identical repeating sections called \textit{unit cells}.

One of the most important properties of materials is its electronic structure, and in particular, its \textit{electronic band structure}. The electronic band structure of a material denotes the various energy levels, as a function of wave vector, that electrons in the material may have. The theory of band structures has formed the basis for understanding electronic and optical properties of materials, as well as the development of electronic devices such as semiconductors \cite{kittel_introduction_2005}. For many simple periodic materials, the tight-binding approximation along with Bloch's theorem and a suitable diagonalization procedure (see Sec.~\ref{TB}) is sufficient to compute the band structure. For more complex materials, more advanced methods such as density functional theory \cite{jones_density_2015} become useful. 

In this project, we study one-dimensional periodic materials using the tight-binding method. Here, the one-dimensional restriction means that the material repeats itself along one direction; however, the atomic arrangement in each individual unit cell may themselves span two or three dimensions. We show how an arbitrary unit cell can be represented as a graph, which is a mathematical structure describing the relationship between objects (see Sec.~\ref{graph}). This representation can then be used to calculate the band structure as well as other properties such as the Fermi level or band gap of real materials, or to theoretically examine relationships between arbitrary unit cell structures and the electronic properties of the corresponding material. In this study, we limit ourselves to non-interacting systems at zero temperature. While graph theory was considered by Bradlyn et. al~\cite{bradlyn_topological_2017, bradlyn_band_2018} as a way to enumerate constraints imposed by crystal symmetries, in this project we provide a more simple and direct analogy between the fields. Overall, graph theory is a powerful computational tool whose usage in solid-state physics is deserving of further research and utilization.

The paper is organized as follows: In Sec.~\ref{background}, we provide background information on the tight-binding method as well as basic graph theory definitions that are sufficient to understand the results of this paper. In Sec.~\ref{analytical_section}, we discuss the connection between graph theory and band theory, and in Sec.~\ref{numerical_section}, we present numerical results for various unit cell structures.

\section{Background}
\label{background}

\subsection{Tight-Binding Method}
\label{TB}

In this section, we outline the \textit{tight-binding method} used to compute the band structures of one-dimensional periodic crystals. The material is taken from Ref.~\cite{hilke_phys_2024}.

The central idea behind the tight-binding method is to approximate the wave function on a lattice structure to be a linear combination of localized orbitals $\ket{n} \approx \phi(\vec{r} - \vec{r}_n)$ to each atom $n$:

\begin{equation}
\psi = \sum_n \psi_n \ket{n}.
\label{LCAO}
\end{equation}

Applying the time-independent Schr\"odinger equation for Hamiltonian $H$ which describes the system, and multiplying from the left by $\bra{m}$ (which represents a localized orbital around atom $m$) we have

\begin{eqnarray}
H \psi &=& E \psi \nonumber \\
\Rightarrow \sum_n \psi_n H \ket{n} &=& E \sum_n \psi_n \ket{n} \nonumber \\
\Rightarrow \sum_n \psi_n \braket{m|H|n} &=& E \sum_n \psi_n \braket{m|n}.
\label{TB_derivation}
\end{eqnarray}

Now, we apply the assumption that each orbital $\ket{n}$ is approximately localized around atom $n$, meaning there is approximately zero overlap between the orbitals around atoms $n$ and $m$. This means that $\{\ket{n}\}$, the set of all localized orbitals, represents an orthonormal basis:

\begin{equation}
\braket{m|n} = \int d^3 r \phi^*(\vec{r} - \vec{r}_m) \phi(\vec{r} - \vec{r}_n) \approx \delta_{m,n}.
\label{TB_approx}
\end{equation}

Eq.~\eqref{TB_approx} is called the \textit{tight-binding approximation} and it holds very well for most systems \cite{hilke_phys_2024}. Additionally, we define

\begin{equation}
\braket{m|H|n} \equiv t_{m,n}
\label{hopping}
\end{equation}

to be the tunneling or hopping amplitude between sites $m$ and $n$. This represents the probability that an electron can tunnel between atoms $m$ and $n$. Since $H$ is Hermitian, we have $|t_{m,n}| = |t_{n,m}|$, as physically expected. When $m = n$, we call $t_{n,n} \equiv v_n$ the self-energy of the atom. Substituting Eq.~\eqref{TB_approx} and Eq.~\eqref{hopping} into Eq.~\eqref{TB_derivation}, we have

\begin{eqnarray}
\sum_n t_{m,n} \psi_n &=& E \sum_n \psi_n \delta_{m,n} \nonumber \\
&=& E \psi_m.
\label{TB_eqn}
\end{eqnarray}

Eq.~\eqref{TB_eqn} is called the \textit{tight-binding equation}. It defines an eigenvalue equation and reduces the problem of finding the allowed energy levels $E$ of the electrons to the problem of diagonalizing the matrix $t_{m,n}$ that defines the hopping amplitudes between the atoms and the self-energies of the atoms.

Materials may have extremely large numbers of atoms. We would like to exploit the periodic nature of many materials to reduce the complexity of the problem. This is done through \textit{Bloch's theorem}, which states that for a periodic material with potential $V$ that has periodicity $a$, i.e. $V(x+a) = V(x)$, the electronic wavefunctions satisfy

\begin{equation}
\psi(x+a) = e^{ika} \psi(x).
\label{Bloch}
\end{equation}

Therefore, the wavefunction of a particle in a given unit cell can be related to that in an adjacent unit cell, and the electronic structure problem can be solved by considering only a single unit cell.

\subsection{Graph Theory}
\label{graph}

In this section, we outline some relevant basic definitions in graph theory. We point the interested reader to a reference such as \cite{wilson_introduction_2009} for more information.

\begin{figure}[t]
    \includegraphics*[width=0.90\columnwidth]{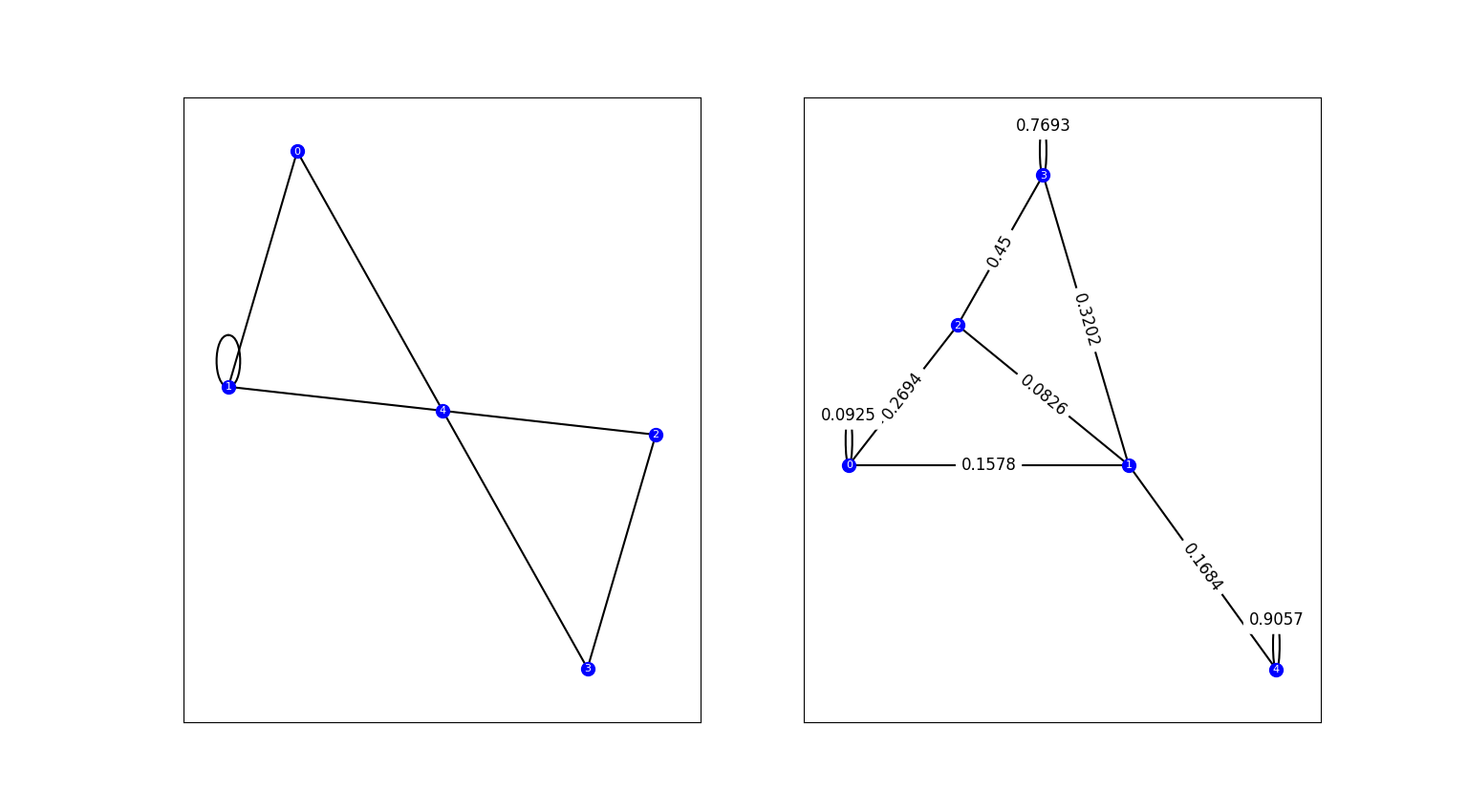}
\caption{Left: example of a simple graph with 5 nodes and 8 edges, one of which is a loop. Right: example of a weighted graph with 5 nodes.}
\label{Figure_1}
\end{figure}

A \textit{graph} is a mathematical structure consisting of a network of nodes (also known as vertices) that are connected by lines called edges. More formally, a simple graph $G$ is a pair $G = (V, E)$ where $V$ is a set whose elements are called vertices and $E$ is a set of unordered pairs $\{v_1, v_2\}$ where $v_1, v_2 \in V$. The elements of $E$ are called edges and represent pairwise relationships between nodes of the graph. A loop is an element of $E$ of the form $\{v_1, v_1\}$, i.e. an edge connecting a node to itself. A simple example of a graph is shown in Fig.~\ref{Figure_1} on the left.

A graph is called connected if it cannot be written as the union of two graphs. Intuitively, this means that there are no isolated vertices or subgraphs in the graph; a path through the graph may be formed between any two vertices.

An important alternative representation of a graph is called its \textit{adjacency matrix}. If the $N$ nodes of a graph are labeled from 1 to $N$, then the $ij$-th entry of the adjacency matrix $a_{ij}$ is defined to be 1 if $\{v_i, v_j\} \in E$ and 0 otherwise. Clearly, the adjacency matrix for a simple graph must be symmetric: $\{v_i, v_j\} \in E \Rightarrow \{v_j, v_i\} \in E$ because the pairs are unordered.

A directed graph (or digraph) is a graph whose edge set $E$ consists of ordered pairs rather than unordered. This means that the edges between nodes may have a specified direction from one node to the other. Further, a weighted graph is a graph in which a number called a weight is assigned to each edge. An example of a weighted graph can be seen in Fig.~\ref{Figure_1} on the right. In the case of a weighted graph, the entries of the adjacency matrix take on the values of the weights rather than 1. The existence of directed edges may break the symmetry of the adjacency matrix. Finally, the eigenvalues of the adjacency matrix are often called the spectrum of the graph.

\section{Calculating Band Structures with Graphs}
\label{analytical_section}

In this section, we explain how graph theory can be used along with the tight-binding method to model periodic chains with arbitrary unit cells and compute their electronic band structure.

Recall from Sec.~\ref{TB} that the problem of electronic band structure can be solved by considering only a single unit cell of the material. Further, to find the energy bands, we must specify the $t_{m,n}$ matrix and compute its eigenvalues. In this project, we use a graphical structure to represent a unit cell of a material that is periodic in one dimension. The representation is defined using the following procedure.\\

\textbf{Procedure 1: Representing periodic materials with graphs.}

\begin{enumerate}
    \item Determine the two outer atoms which connect a unit cell to adjacent cells. Label the leftmost atom as atom 1 and the rightmost as atom $m$.
    \item Label the remaining atoms in the unit cell from 2 to $N$, where $m \leq N$ is not included.
    \item Define the node set of the graph as a one-to-one correspondence to the set of atoms in the unit cell: $V = \{v_1, \ldots, v_N\}$.
    \item For each tunneling connection between atoms $i$ and $j$ within the unit cell, define an edge $\{v_i, v_j\}$ in the edge set of the graph. These are all undirected edges.
    \item Define the weight of the edge $\{v_i, v_j\}$ connecting nodes $i$ and $j$ to be the tunneling amplitude between atoms $i$ and $j$: $w(ij) = t_{i,j}$. 
    \item For any atoms with nonzero on-site energies, define a loop on the corresponding graph node with weight corresponding to the on-site energy of that atom: $w_{ii} = v_i$.
    \item Introduce an additional directed edge from node $m$ to node 1 with weight $e^{ika}$ where $a$ is the periodicity of the material.
    \item Introduce an additional directed edge from node 1 to node $m$ with weight $e^{-ika}$ where $a$ is the periodicity of the material.
\end{enumerate} 

The adjacency matrix of the resulting graph will then, by construction, be equal to the matrix $t_{m,n}$, whose diagonalization then gives the energy bands of the material. In general, the graph is a weighted digraph. However, before step 7, it is a weighted simple graph, as only the edges defined in steps 7 and 8 are directed. These two entries of the adjacency matrix break its symmetry and instead make the matrix Hermitian as the weights are complex conjugates of one another. Further, the graph must be connected. We limit ourselves to the case where adjacent unit cells are connected by a single atom (atom 1 on the left and atom $m$ on the right).

The main benefit of this representation is that it can be used to represent any arbitrary unit cell: It may have an arbitrary shape and number of atoms as well an arbitrary set of tunneling amplitudes and set of on-site energies. The graphical representation then provides an efficient and generalizable method for numerically storing these configurations. Further, results from spectral graph theory (see, for example, Ref.~\cite{chung_spectral_1997}) on the adjacency matrix of the graph immediately apply to the energy bands of the material.

\section{Numerical Results}
\label{numerical_section}

For this project, we provide a simple algorithm to compute the band structure of a material given an input graph representing a unit cell of the atom. In addition, we provide an algorithm to compute the Fermi level and band gap of a material for various electron fillings. We adopt the definition that the Fermi level is equal to the chemical potential at zero temperature. The code for the algorithms can be found on \href{https://github.com/robertgerstner/BandStructureGraphs}{GitHub} \cite{gerstner_bandstructuregraphs_2024}. We used the NetworkX Python package \cite{hagberg_exploring_2008} to work with graphs.

In this section of the paper, we outline some numerical results which exemplify the usage of this method. We start with simple circular chains of arbitrary numbers of atoms and then proceed with more complex random graphs. Note that in all presented graph drawings, the displayed graph represents a single unit cell of the periodic material.

\subsection{Circular Chains}

A circular chain of atoms is a lattice in which each unit cell contains $N$ atoms connected with nearest-neighbour hopping and periodic boundary conditions.

\begin{figure}[t]
    \includegraphics*[width=0.99\columnwidth]{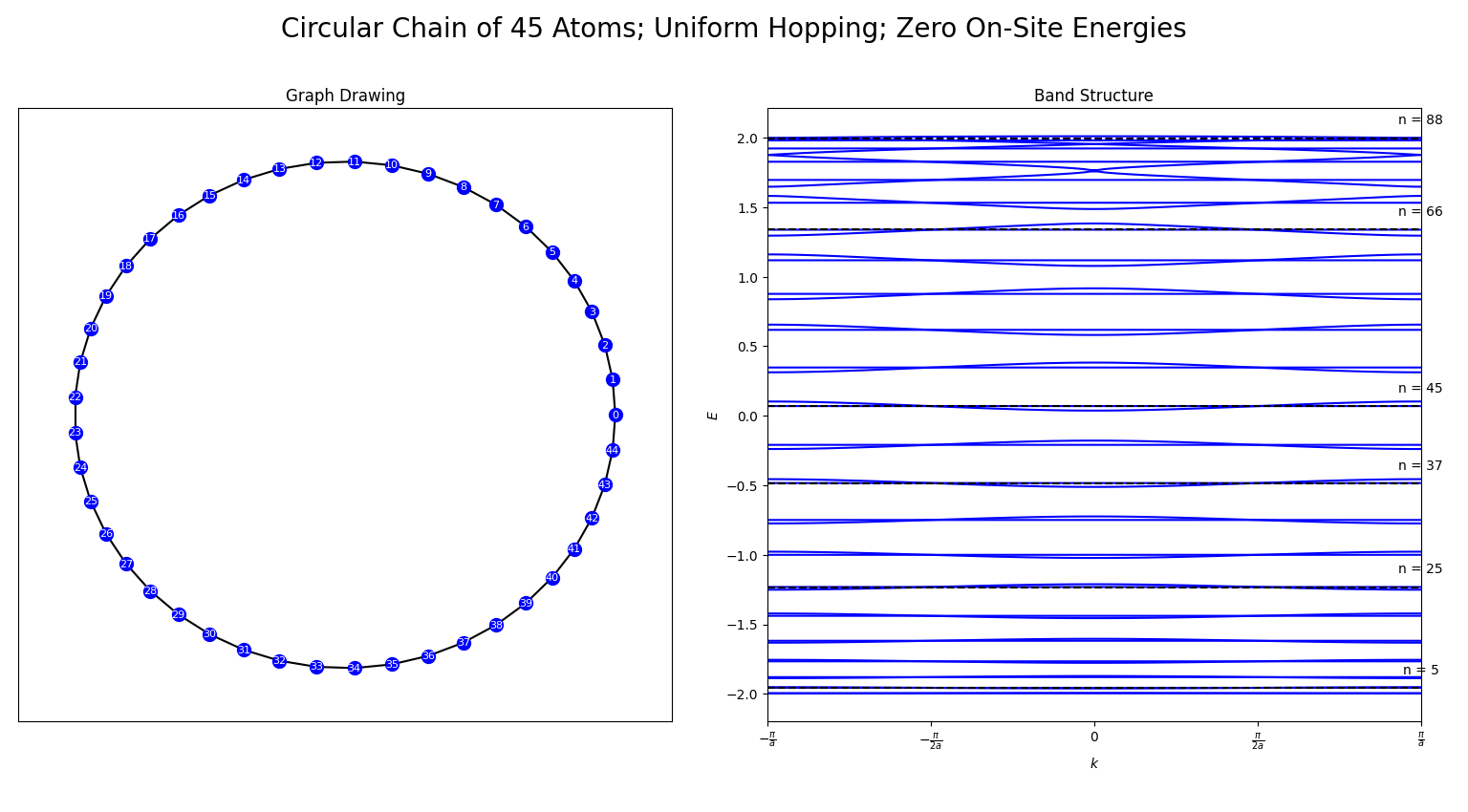}
\caption{Left: Graph drawing of the unit cell of a circular chain of 45 atoms with unit hoppings and zero on-site energies. In white is the node labeling. Right: Computed band structure of the material through the Brillouin zone. In dashed black lines are the Fermi levels for various electron fillings, where $n$ represents the number of electrons in each unit cell.}
\label{Figure_2}
\end{figure}

First, we consider the simplest case of uniform hopping amplitudes and zero on-site energies. We show the results, including the graph drawing, the band structure calculation, and various Fermi level calculations, for a relatively large number of $N = 45$ atoms per unit cell in Fig.~\ref{Figure_2}.

\begin{figure}[t]
    \includegraphics*[width=0.90\columnwidth]{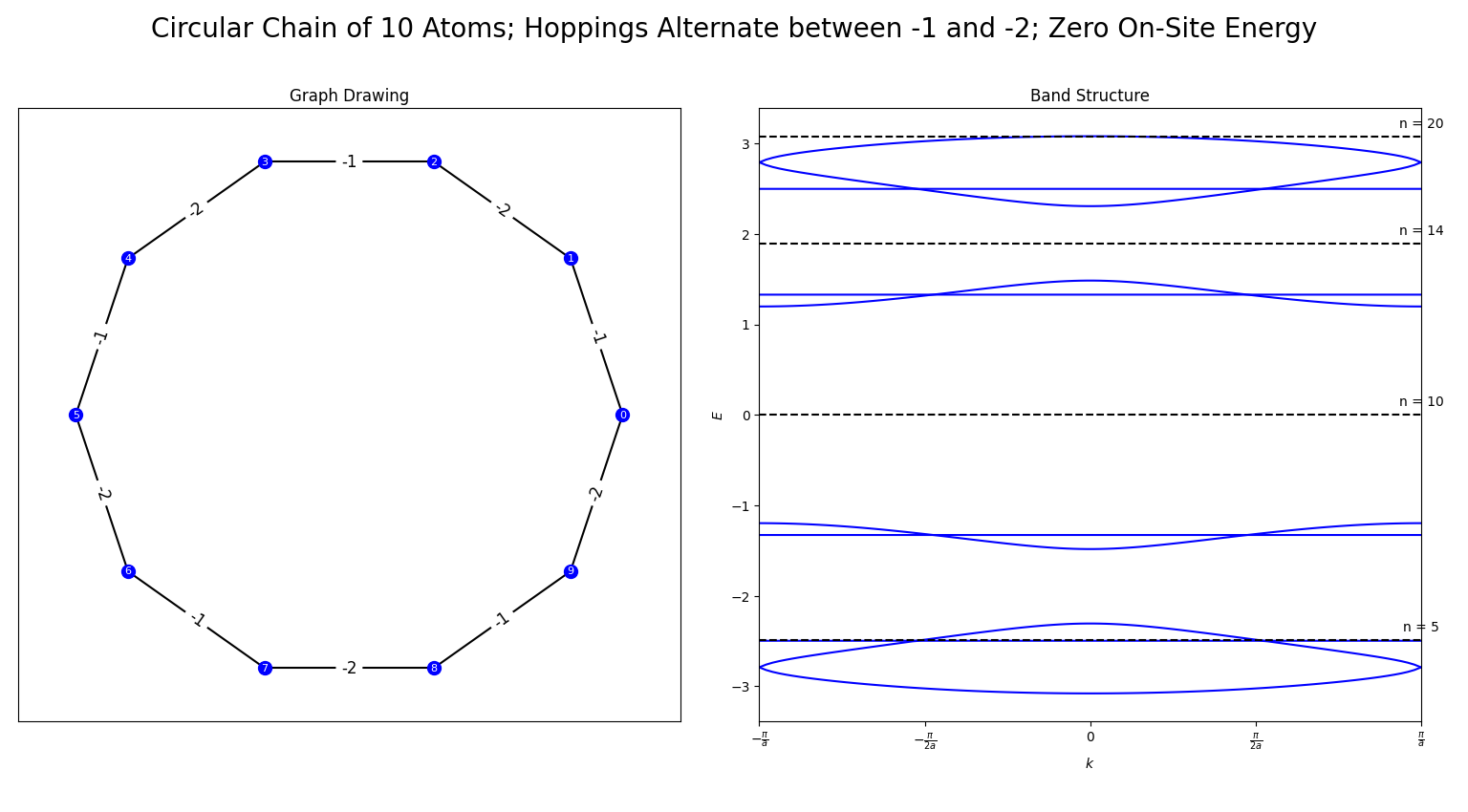}
\caption{Graph drawing and band structure for a circular chain of 10 atoms with hoppings alternating between -1 and -2 and zero on-site energies.}
\label{Figure_3}
\end{figure}

\begin{figure}[t]
    \includegraphics*[width=0.90\columnwidth]{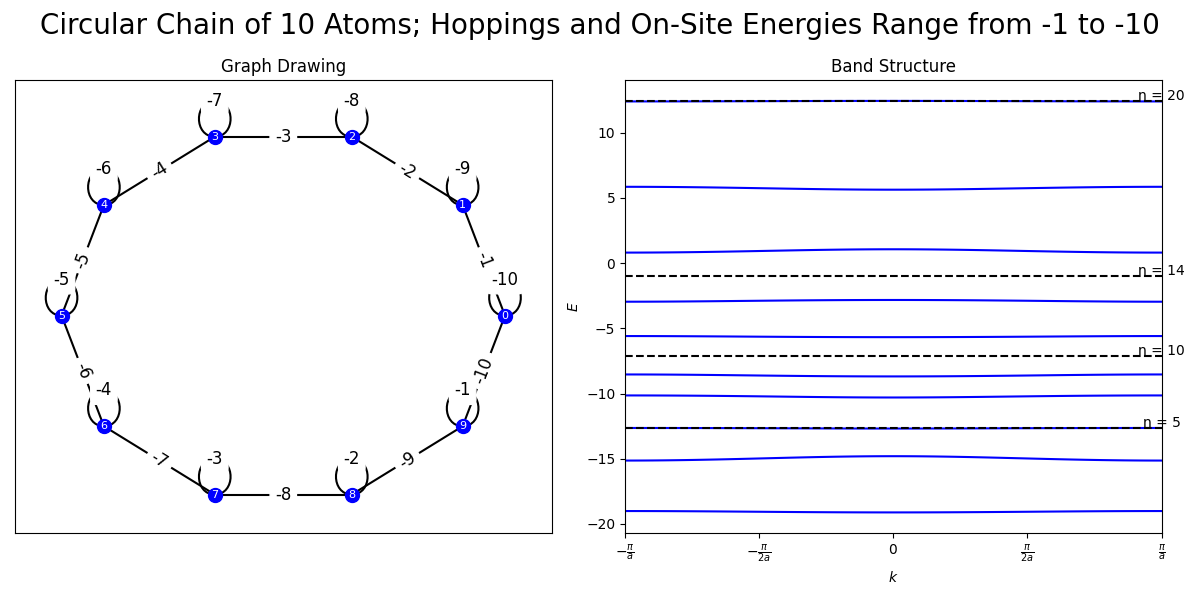}
\caption{Graph drawing and band structure for a circular chain of 10 atoms with hoppings and on-site energies ranging from $-1$ to $-10$.}
\label{Figure_4}
\end{figure}

\begin{figure}[t]
    \includegraphics*[width=0.90\columnwidth]{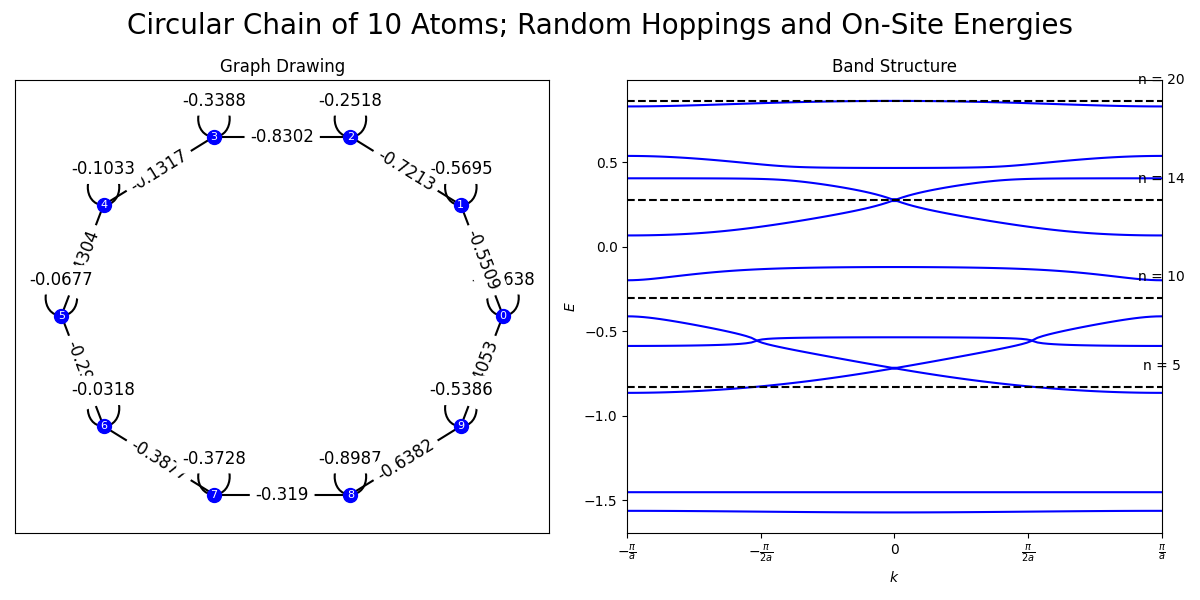}
\caption{Graph drawing and band structure for a circular chain of 10 atoms with hoppings and on-site energies drawn randomly between 0 and $-1$.}
\label{Figure_5}
\end{figure}

We now consider a few more interesting cases of circular chains. In Fig.~\ref{Figure_3}, we consider hoppings alternating betwen $-1$ and $-2$. We find that the band structure has similar shape but with a much larger band gap for electron half-filling. In Fig.~\ref{Figure_4}, we consider linearly increasing hopping amplitudes and on-site energies. We find that the bands become nearly flat in this case. This is a notable result as flat bands are of great interest in current condensed matter physics due to their interesting topological properties \cite{checkelsky_flat_2024}.  Finally, in Fig.~\ref{Figure_5}, we consider hoppings and on-site energies chosen at random in the range $[-1,0]$. 

\subsection{Random Structures}

The true power of this method is revealed in the ability to represent unit cells of arbitrary complexity. We demonstrate this through the calculation of band structures of completely random unit cell structures. We generate random unit cell graphs via their adjacency matrix using the following procedure:\\

\textbf{Procedure 2: Random unit cell generation.}
\begin{enumerate}
\item Define an $N\times N$ matrix with element magnitudes randomly chosen from the range $[0,1]$ with some density $d$, $0 < d \leq 1$. Here, $d$ represents the percentage of matrix elements that are nonzero.
\item To ensure the matrix is symmetric, extract the upper triangular portion including the diagonal. Then, extract the upper triangular portion not including the diagonal, take the transpose, and add it to the previously-extracted upper-triangular matrix.
\item Check if the resulting adjacency matrix corresponds to a connected graph. If not, repeat the process. 
\item Insert terms of the form $e^{ika}$ and $e^{-ika}$ into corresponding entries to satisfy Bloch's theorem.
\item Compute the band structure, Fermi level and/or band gap.
\end{enumerate}

\begin{figure}[t]
    \includegraphics*[width=0.95\columnwidth]{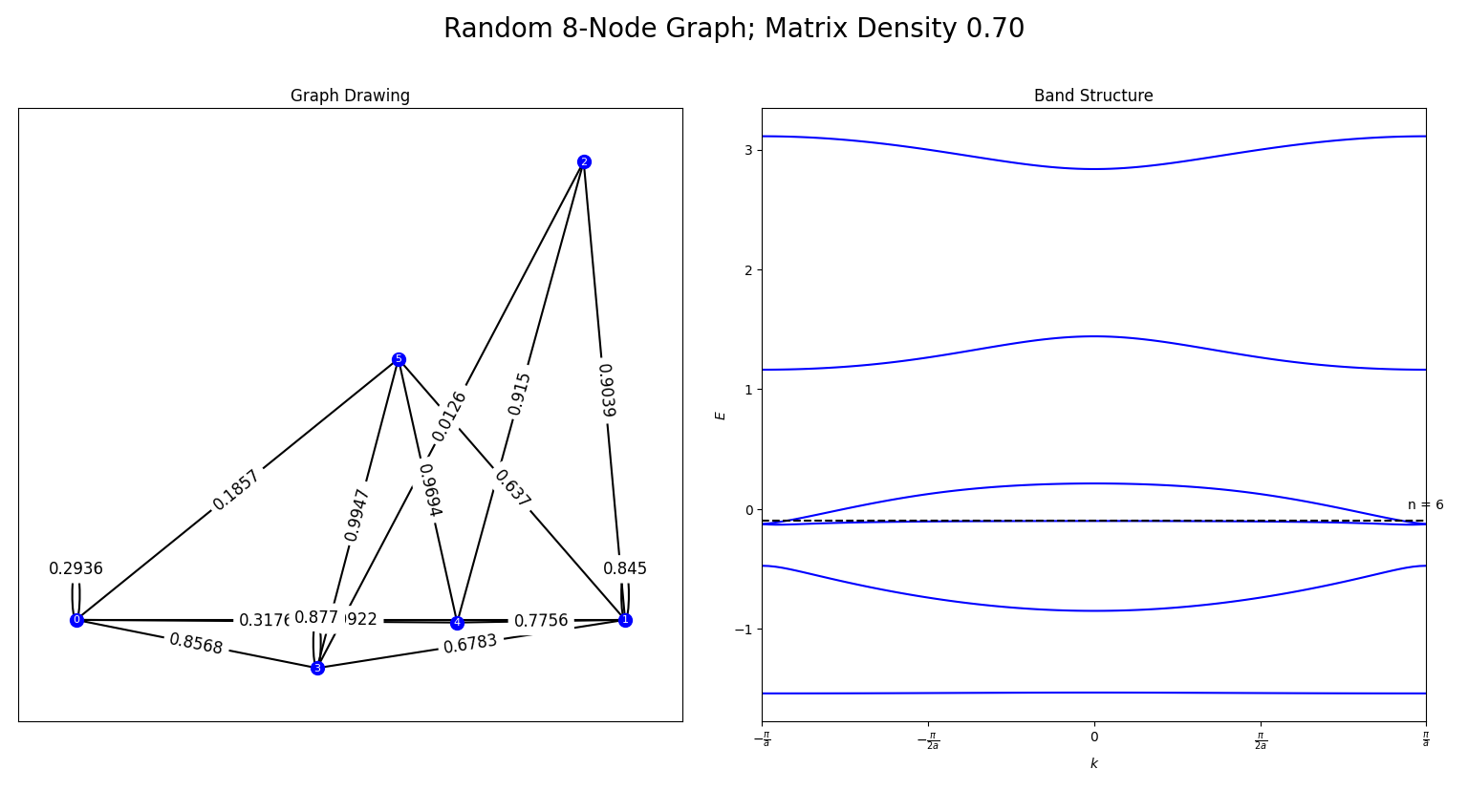}
\caption{Graph drawing and band structure for a random chain of 6 atoms with matrix density 0.70. At half filling, the Fermi level lies in a band, and so this hypothetical material would have nonzero conductivity.}
\label{Figure_6}
\end{figure}

\begin{figure}[t]
    \includegraphics*[width=0.95\columnwidth]{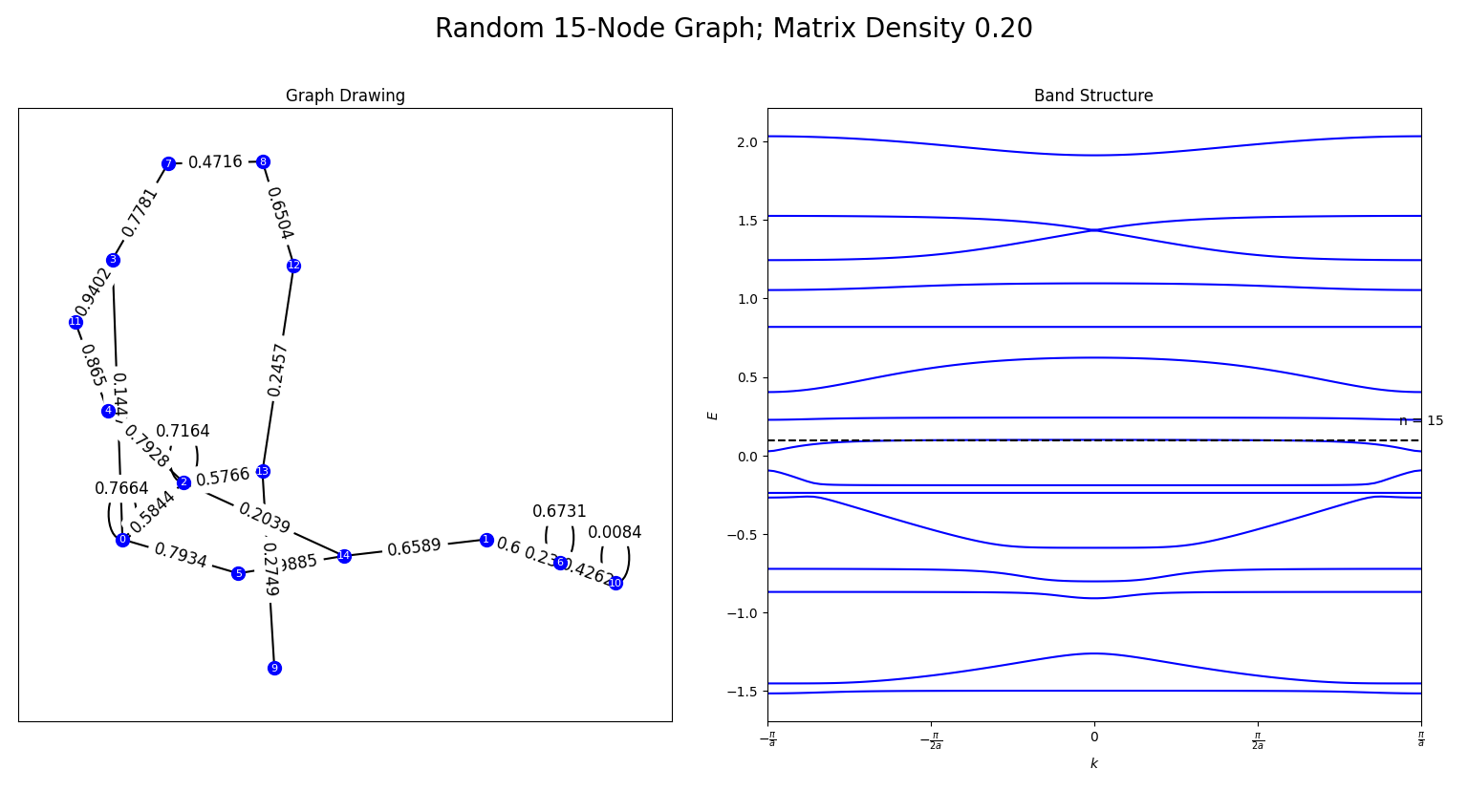}
\caption{Graph drawing and band structure for a random chain of 15 atoms with matrix density 0.20. At half filling, the Fermi level lies in a band, and so this hypothetical material would have nonzero conductivity.}
\label{Figure_7}
\end{figure}

\begin{figure}[H]
    \includegraphics*[width=0.99\columnwidth]{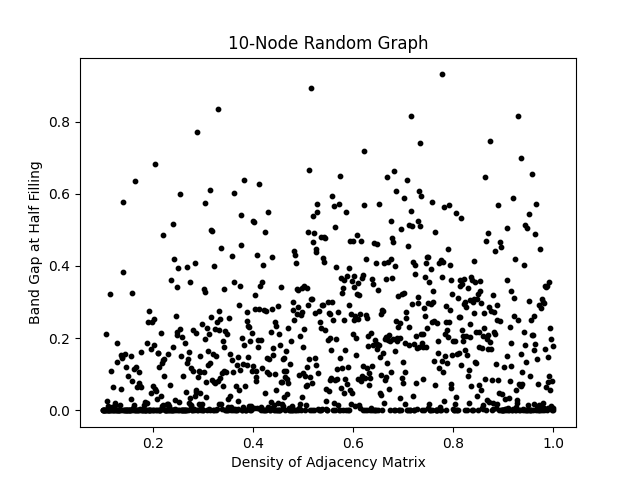}
\caption{Relation between the band gap of a hypothetical material at half filling and the density of its associated adjacency matrix. 1000 random 10-node graphs were generated with $0.1 \leq d \leq 1$ and used as data points.}
\label{Figure_8}
\end{figure}

We exemplify this procedure in Figs.~\ref{Figure_6} and ~\ref{Figure_7} for $N = 6$ and $N = 15$ respectively. Using the SciPy module, we can also specify the sparsity of the generated adjacency matrix, which controls how many pairwise connections are in the unit cell. In particular, the density $d$ specifies the fraction of matrix elements that are nonzero. For the smaller chain of 6 sites in Fig.~\ref{Figure_7} we chose a density of 0.70, and for the larger chain of 15 sites in Fig.~\ref{Figure_8} we chose a density of 0.20. Using the band structures, we can determine whether the material is conducting or insulating at half filling based on whether the Fermi level lies in a band gap.

To conclude this section, we use the graphical representation to study the relationship between the connectivity of a material's unit cell and whether it conducts or insulates at zero temperature and half filling. Here, half filling means one electron per atom contributes to the filling of energy levels. The connectivity of a unit cell can be represented by the density of its associated adjacency matrix. We collected data for 1000 randomly-generated unit cells of 10 atoms with matrix densities ranging from 0.10 to 1.00 and found the band gap at half filling for each. In this case, a band gap of zero indicates conductivity at zero temperature, while a nonzero band gap indicates insulation. The results are shown in Fig.~\ref{Figure_8}. We find that there is no relationship between the connectivity of the unit cell and its band gap. \\\\

\section{Conclusion}
\label{conclusion}

We showed that graphical structures can be used to effectively represent unit cells of one-dimensional periodic materials and to compute and study the associated band structures. The method permits unit cells of arbitrary structure, which may benefit in the study of theoretical relationships in band structures or for studying novel complex materials. Further research could involve applying theorems from spectral graph theory to determine more efficient diagonalization methods or to reveal deeper connections between the fields. Additionally, the versatility of graphs makes generalization to two- or three-dimensional periodic materials possible. The method could also be applied to computing phonon dispersion relations under the harmonic approximation \cite{hilke_phys_2024}. Overall, this paper provides an introduction to the applicability of graph theory as a computational tool for solid-state physics.

\bibliography{references}

\begin{thebibliography}{10}%
\makeatletter
\providecommand \@ifxundefined [1]{%
 \@ifx{#1\undefined}
}%
\providecommand \@ifnum [1]{%
 \ifnum #1\expandafter \@firstoftwo
 \else \expandafter \@secondoftwo
 \fi
}%
\providecommand \@ifx [1]{%
 \ifx #1\expandafter \@firstoftwo
 \else \expandafter \@secondoftwo
 \fi
}%
\providecommand \natexlab [1]{#1}%
\providecommand \enquote  [1]{``#1''}%
\providecommand \bibnamefont  [1]{#1}%
\providecommand \bibfnamefont [1]{#1}%
\providecommand \citenamefont [1]{#1}%
\providecommand \href@noop [0]{\@secondoftwo}%
\providecommand \href [0]{\begingroup \@sanitize@url \@href}%
\providecommand \@href[1]{\@@startlink{#1}\@@href}%
\providecommand \@@href[1]{\endgroup#1\@@endlink}%
\providecommand \@sanitize@url [0]{\catcode `\\12\catcode `\$12\catcode `\&12\catcode `\#12\catcode `\^12\catcode `\_12\catcode `\%12\relax}%
\providecommand \@@startlink[1]{}%
\providecommand \@@endlink[0]{}%
\providecommand \url  [0]{\begingroup\@sanitize@url \@url }%
\providecommand \@url [1]{\endgroup\@href {#1}{\urlprefix }}%
\providecommand \urlprefix  [0]{URL }%
\providecommand \Eprint [0]{\href }%
\providecommand \doibase [0]{https://doi.org/}%
\providecommand \selectlanguage [0]{\@gobble}%
\providecommand \bibinfo  [0]{\@secondoftwo}%
\providecommand \bibfield  [0]{\@secondoftwo}%
\providecommand \translation [1]{[#1]}%
\providecommand \BibitemOpen [0]{}%
\providecommand \bibitemStop [0]{}%
\providecommand \bibitemNoStop [0]{.\EOS\space}%
\providecommand \EOS [0]{\spacefactor3000\relax}%
\providecommand \BibitemShut  [1]{\csname bibitem#1\endcsname}%
\let\auto@bib@innerbib\@empty
\bibitem [{\citenamefont {Kittel}(2005)}]{kittel_introduction_2005}%
  \BibitemOpen
  \bibfield  {author} {\bibinfo {author} {\bibfnamefont {C.}~\bibnamefont {Kittel}},\ }\href@noop {} {\emph {\bibinfo {title} {Introduction to solid state physics}}},\ \bibinfo {edition} {8th}\ ed.\ (\bibinfo  {publisher} {Wiley},\ \bibinfo {address} {Hoboken, NJ},\ \bibinfo {year} {2005})\BibitemShut {NoStop}%
\bibitem [{\citenamefont {Jones}(2015)}]{jones_density_2015}%
  \BibitemOpen
  \bibfield  {author} {\bibinfo {author} {\bibfnamefont {R.}~\bibnamefont {Jones}},\ }\bibfield  {title} {\bibinfo {title} {Density functional theory: {Its} origins, rise to prominence, and future},\ }\href {https://doi.org/10.1103/RevModPhys.87.897} {\bibfield  {journal} {\bibinfo  {journal} {Reviews of Modern Physics}\ }\textbf {\bibinfo {volume} {87}},\ \bibinfo {pages} {897} (\bibinfo {year} {2015})}\BibitemShut {NoStop}%
\bibitem [{\citenamefont {Bradlyn}\ \emph {et~al.}(2017)\citenamefont {Bradlyn}, \citenamefont {Elcoro}, \citenamefont {Cano}, \citenamefont {Vergniory}, \citenamefont {Wang}, \citenamefont {Felser}, \citenamefont {Aroyo},\ and\ \citenamefont {Bernevig}}]{bradlyn_topological_2017}%
  \BibitemOpen
  \bibfield  {author} {\bibinfo {author} {\bibfnamefont {B.}~\bibnamefont {Bradlyn}}, \bibinfo {author} {\bibfnamefont {L.}~\bibnamefont {Elcoro}}, \bibinfo {author} {\bibfnamefont {J.}~\bibnamefont {Cano}}, \bibinfo {author} {\bibfnamefont {M.~G.}\ \bibnamefont {Vergniory}}, \bibinfo {author} {\bibfnamefont {Z.}~\bibnamefont {Wang}}, \bibinfo {author} {\bibfnamefont {C.}~\bibnamefont {Felser}}, \bibinfo {author} {\bibfnamefont {M.~I.}\ \bibnamefont {Aroyo}},\ and\ \bibinfo {author} {\bibfnamefont {B.~A.}\ \bibnamefont {Bernevig}},\ }\bibfield  {title} {\bibinfo {title} {Topological quantum chemistry},\ }\href {https://doi.org/10.1038/nature23268} {\bibfield  {journal} {\bibinfo  {journal} {Nature}\ }\textbf {\bibinfo {volume} {547}},\ \bibinfo {pages} {298} (\bibinfo {year} {2017})}\BibitemShut {NoStop}%
\bibitem [{\citenamefont {Bradlyn}\ \emph {et~al.}(2018)\citenamefont {Bradlyn}, \citenamefont {Elcoro}, \citenamefont {Vergniory}, \citenamefont {Cano}, \citenamefont {Wang}, \citenamefont {Felser}, \citenamefont {Aroyo},\ and\ \citenamefont {Bernevig}}]{bradlyn_band_2018}%
  \BibitemOpen
  \bibfield  {author} {\bibinfo {author} {\bibfnamefont {B.}~\bibnamefont {Bradlyn}}, \bibinfo {author} {\bibfnamefont {L.}~\bibnamefont {Elcoro}}, \bibinfo {author} {\bibfnamefont {M.~G.}\ \bibnamefont {Vergniory}}, \bibinfo {author} {\bibfnamefont {J.}~\bibnamefont {Cano}}, \bibinfo {author} {\bibfnamefont {Z.}~\bibnamefont {Wang}}, \bibinfo {author} {\bibfnamefont {C.}~\bibnamefont {Felser}}, \bibinfo {author} {\bibfnamefont {M.~I.}\ \bibnamefont {Aroyo}},\ and\ \bibinfo {author} {\bibfnamefont {B.~A.}\ \bibnamefont {Bernevig}},\ }\bibfield  {title} {\bibinfo {title} {Band connectivity for topological quantum chemistry: {Band} structures as a graph theory problem},\ }\href {https://doi.org/10.1103/PhysRevB.97.035138} {\bibfield  {journal} {\bibinfo  {journal} {Physical Review B}\ }\textbf {\bibinfo {volume} {97}},\ \bibinfo {pages} {035138} (\bibinfo {year} {2018})}\BibitemShut {NoStop}%
\bibitem [{\citenamefont {Hilke}(2024)}]{hilke_phys_2024}%
  \BibitemOpen
  \bibfield  {author} {\bibinfo {author} {\bibfnamefont {M.}~\bibnamefont {Hilke}},\ }\href {https://www.physics.mcgill.ca/~hilke/558/558.html} {\bibinfo {title} {Phys 558: {Solid} {State} {Physics}}} (\bibinfo {year} {2024})\BibitemShut {NoStop}%
\bibitem [{\citenamefont {Wilson}(2009)}]{wilson_introduction_2009}%
  \BibitemOpen
  \bibfield  {author} {\bibinfo {author} {\bibfnamefont {R.~J.}\ \bibnamefont {Wilson}},\ }\href@noop {} {\emph {\bibinfo {title} {Introduction to graph theory}}},\ \bibinfo {edition} {4th}\ ed.\ (\bibinfo  {publisher} {Prentice Hall},\ \bibinfo {address} {Harlow Munich},\ \bibinfo {year} {2009})\BibitemShut {NoStop}%
\bibitem [{\citenamefont {Chung}(1997)}]{chung_spectral_1997}%
  \BibitemOpen
  \bibfield  {author} {\bibinfo {author} {\bibfnamefont {F.~R.~K.}\ \bibnamefont {Chung}},\ }\href@noop {} {\emph {\bibinfo {title} {Spectral graph theory}}},\ \bibinfo {series} {Regional conference series in mathematics}\ No.\ \bibinfo {number} {no. 92}\ (\bibinfo  {publisher} {Published for the Conference Board of the mathematical sciences by the American Mathematical Society},\ \bibinfo {address} {Providence, R.I},\ \bibinfo {year} {1997})\BibitemShut {NoStop}%
\bibitem [{\citenamefont {Gerstner}(2024)}]{gerstner_bandstructuregraphs_2024}%
  \BibitemOpen
  \bibfield  {author} {\bibinfo {author} {\bibfnamefont {R.}~\bibnamefont {Gerstner}},\ }\href {https://github.com/robertgerstner/BandStructureGraphs/tree/main} {\bibinfo {title} {{BandStructureGraphs}}} (\bibinfo {year} {2024})\BibitemShut {NoStop}%
\bibitem [{\citenamefont {Hagberg}\ \emph {et~al.}(2008)\citenamefont {Hagberg}, \citenamefont {Schult},\ and\ \citenamefont {Swart}}]{hagberg_exploring_2008}%
  \BibitemOpen
  \bibfield  {author} {\bibinfo {author} {\bibfnamefont {A.~A.}\ \bibnamefont {Hagberg}}, \bibinfo {author} {\bibfnamefont {D.~A.}\ \bibnamefont {Schult}},\ and\ \bibinfo {author} {\bibfnamefont {P.~J.}\ \bibnamefont {Swart}},\ }\bibfield  {title} {\bibinfo {title} {Exploring {Network} {Structure}, {Dynamics}, and {Function} using {NetworkX}},\ }in\ \href@noop {} {\emph {\bibinfo {booktitle} {Proceedings of the 7th {Python} in {Science} {Conference}}}}\ (\bibinfo {year} {2008})\ pp.\ \bibinfo {pages} {11--16}\BibitemShut {NoStop}%
\bibitem [{\citenamefont {Checkelsky}\ \emph {et~al.}(2024)\citenamefont {Checkelsky}, \citenamefont {Bernevig}, \citenamefont {Coleman}, \citenamefont {Si},\ and\ \citenamefont {Paschen}}]{checkelsky_flat_2024}%
  \BibitemOpen
  \bibfield  {author} {\bibinfo {author} {\bibfnamefont {J.~G.}\ \bibnamefont {Checkelsky}}, \bibinfo {author} {\bibfnamefont {B.~A.}\ \bibnamefont {Bernevig}}, \bibinfo {author} {\bibfnamefont {P.}~\bibnamefont {Coleman}}, \bibinfo {author} {\bibfnamefont {Q.}~\bibnamefont {Si}},\ and\ \bibinfo {author} {\bibfnamefont {S.}~\bibnamefont {Paschen}},\ }\bibfield  {title} {\bibinfo {title} {Flat bands, strange metals and the {Kondo} effect},\ }\href {https://doi.org/10.1038/s41578-023-00644-z} {\bibfield  {journal} {\bibinfo  {journal} {Nature Reviews Materials}\ }\textbf {\bibinfo {volume} {9}},\ \bibinfo {pages} {509} (\bibinfo {year} {2024})}\BibitemShut {NoStop}%
\end{thebibliography}%

\end{document}